\documentclass{article}

\pdfoutput=1
\usepackage{arxiv}

\usepackage{textgreek}
\usepackage[utf8]{inputenc} 
\usepackage[T1]{fontenc}    
\usepackage{hyperref}       
\usepackage{url}            
\usepackage{booktabs}       
\usepackage{amsfonts}       
\usepackage{nicefrac}       
\usepackage{microtype}      
\usepackage{lipsum}
\usepackage{graphicx}
\graphicspath{ {./images/} }

\title{Successive training of a generative adversarial network for the design of an optical cloak}

\author{
 André-Pierre Blanchard-Dionne \\
  École Polytechnique Fédérale de Lausanne\\
  Lausanne, Switzerland \\
  \texttt{andre-pierre.blanchard-dionne@epfl.ch} \\
   \And
 Olivier J.F. Martin \\
  École Polytechnique Fédérale de Lausanne\\
  Lausanne, Switzerland \\
  \texttt{olivier.martin@epfl.ch} \\}

\begin{document}
\maketitle
\begin{abstract}
We present an optimization algorithm based on a deep convolution generative adversarial network (DCGAN) to design a 2-Dimensional optical cloak. The optical cloak consists in a shell of uniform and isotropical dielectric material, and the cloaking is achieved via the geometry of the shell. We use a feedback loop from the solutions of the DCGAN to successively retrain it and improve its ability to predict and find optimal geometries.
\end{abstract}


\section{Introduction}
The theoretical idea of an optical cloak was first suggested in 2006 by JF Pendry in a paper of transformation optics \cite{pendry2006controlling} and was later demonstrated experimentally \cite{schurig2006metamaterial}. It introduced the concept that a electromagnetic wavefront could be bent inside a circular shell of carefully chosen dielectric and magnetic materials and remain unchanged upon its exit, thus concealing an object inside of it. The proposed materials of the shell consisted of non-uniform anisotropic dielectric and magnetic materials. Recently, similar cloaking behaviour was found to exist in shell designs of more simple all-dielectric isotropical and uniform material \cite{andkjaer2012towards,andkjaer2011topology,fujii2013level}. The working principle for those shells resides in their geometry which can lead to the bending of the wavefront around the object to hide. The optimal geometries in those reported cases were found using topology optimization methods.

\begin{figure}[htbp]
\centering
\includegraphics[width=120mm]{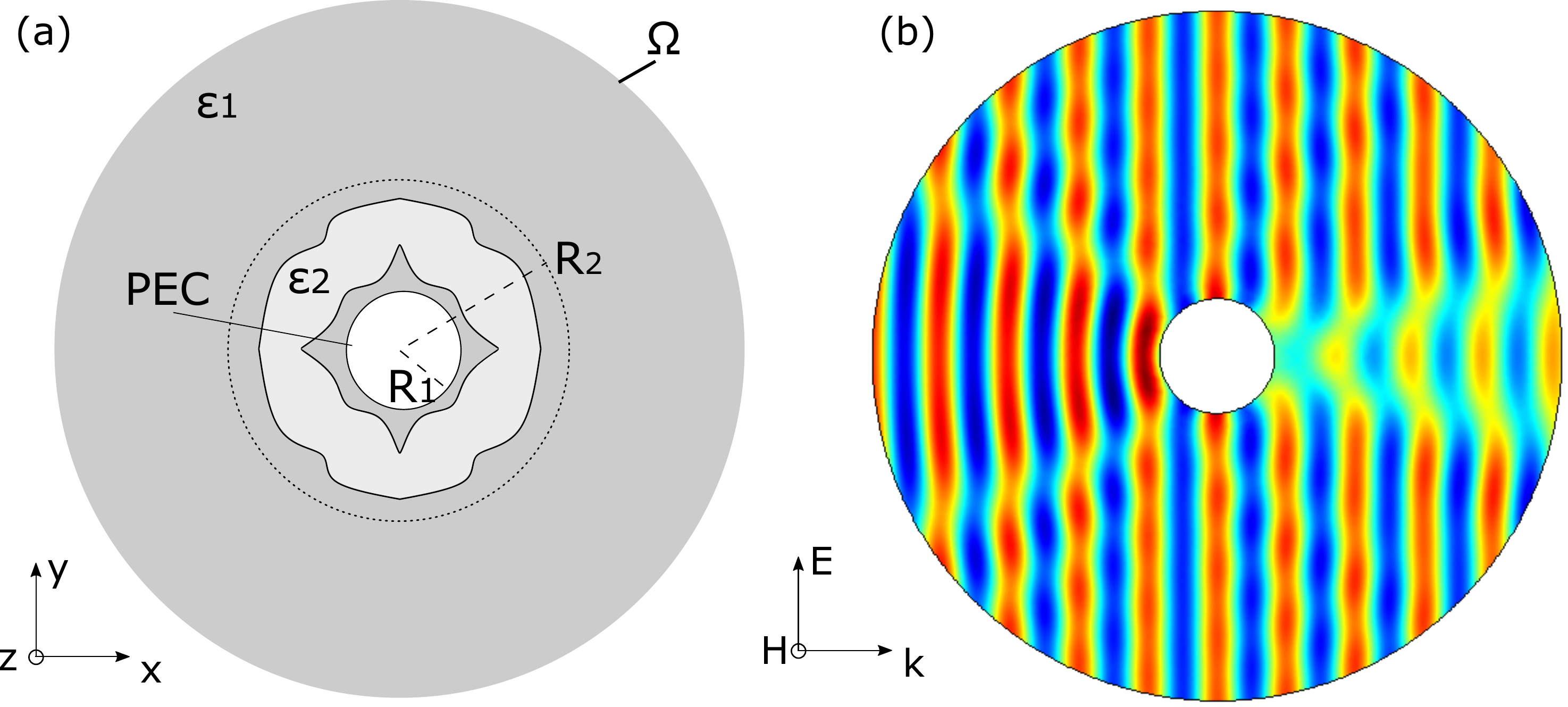}
\caption{(a) Schematic of the simulation domain. The object at r=R\textsubscript{1} is represented by a perfect electrical conductor (PEC). The shell consists of a medium of permittivity $\epsilon_2 = 2$ inside the region $R_1 < r < R_2$. The scattered fields are calculated at the boundary of the domain $\Omega$ (b) Total transverse fields $H_z$ when no cloaks are included in the design.  \label{fig:geo}}
\end{figure}

In this paper we use a deep learning genarative algorithm to accomplish the optimization of the geometry of an all-dielectric isotropic shell for cloaking. Generative networks have recently been demonstrated as useful tools for finding global solutions to inverse engineering problems \cite{sanchez2018inverse} and have found applications in nanophotonics for the design of metasurfaces \cite{liu2018generative,jiang2019simulator,an2020freeform,hodge2019rf,ma2019probabilistic}, nanostructures \cite{so2019designing}, metagratings \cite{jiang2019free} and power splitters \cite{tang2020generative}. We suggest to use a generative adversarial network \cite{goodfellow2014generative} in a feedback loop in order to enable the network to improve itself and find an optimal configuration for cloaking. The procedure is done in several steps: First, several shells geometries are randomly created and are simulated to obtain their scattering coefficient (see Figure \ref{fig:geo}(a)). We then train a forward network (FN) to predict the scattered field amplitude from each geometry. A deep convolutional generative adversarial network (DCGAN) coupled to the FN is then used to generate new solutions which are aimed at minimizing the scattered fields. We then implement the loop to reuse the generated solutions to improve both the CN and the DCGAN for sufficient iterations until a satisfactory solution is found or the generative algorithm doesn't improve anymore.

\section{Methods}
\label{sec:headings}

\subsection{FEM simulations}
In order to obtain the scattering fields amplitude from a given cloak geometry, finite-element simulations using COMSOL's RF module are used. The simulation domain is presented in Figure \ref{fig:geo} and has an exterior circular boundary with a radius of 12 \textmu m. Another circular boundary of perfect electrical conductor (PEC) is located at R\textsubscript{1} =1 \textmu m and constitutes the object to conceal. Multiple polygons of dielectric constant \textepsilon\textsubscript{2} = 2 are included in the region R\textsubscript{1} < r < R\textsubscript{2} to form the cloaking shell. A background illumination electromagnetic field defined by $\vec{E} = E_0 e^{i k_0 x} \hat{y} $ is used and the scattered fields are solved with scattering boundary conditions at \textPsi. We then integrate the relative Poynting vector at the exterior boundary of the simulation to obtain the scattering metric to minimize :

\begin{equation}
    \Psi =\frac{1}{2} \int_{\Omega}{ \overrightarrow{E}_{rel} \times \overrightarrow{H}_{rel}^*} \; dC
\end{equation}

where $\overrightarrow{E}_{rel} = \overrightarrow{E}_{total}-\overrightarrow{E}_{background}$. 

We use an object of radius R\textsubscript{1}=1 \textmu m and a shell of R\textsubscript{2}=3 \textmu m for the simulations, and the wavelength is set at \textlambda = $R_2 / 2.5 $ = 1.2 \textmu m. A total of 13000 simulations were performed with randomly generated shell geometries to form the initial dataset. The geometries were generated using the union of randomly generated curves inside the defined shell radius. We use a shell which is symmetrical in -x and -y according to the symmetry of the problem and also to assure continuity of the shell around the object.

\subsection{Neural networks}

The first part of the model consists in a forward predictive model of the scattering coefficient $\Phi$ as a function of the shell geometry. The input image consists in $64 \times 64$ binary images where the region of dielectric constant $\epsilon_2$ is represented by 1 and $\epsilon_1$ by 0. Since the shells are symmetrical in -x and -y, the images are taken for only one quadrant of the shell. A convolution network is then used, which consists in 4 convolution layers and 2 dense layers that takes a $(64 \times 64)$ image data and creates a single digit output. The network is trained with an Adam optimizer \cite{kingma2014adam} with a learning rate of $1\cdot 10^{-4}$ and the mean squared error is used as the loss function. Precise details of the network are provided in the supplementary information section.

\begin{figure}[htbp]
\centering
\includegraphics[width=80mm]{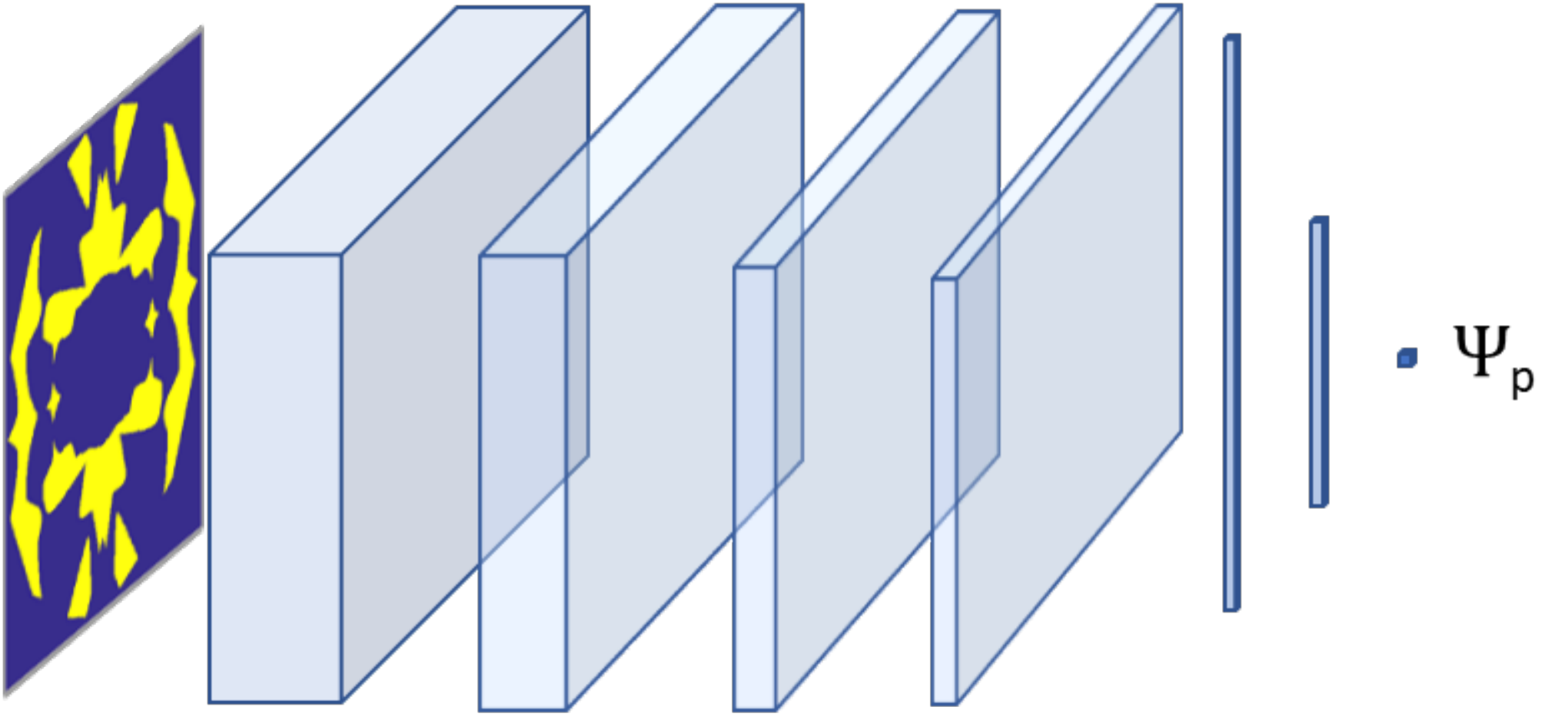}
\caption{Forward predictive network. The input is the image of a cloaking shell configuration. The network consists in 4 convolution layers and 2 dense layers, and the output is the scattering coefficient $\Psi$ \label{fig:forward_model}}
\end{figure}

The generative adversarial network consists of two different networks, a generator and a discriminator, which work in opposition to create new shell geometries similar to those of the dataset. The generator is a transposed convolution network, which takes a random noise vector of dimension 200 as input and creates a new "fake" image of a cloaking shell as output. It consists of four layers of transposed convolution layers (Conv2DTranspose), which upsamples a matrix by applying strided convolution operations. The discriminator is a convolution network and takes a set of "fake" images (from the generator) and real images (from the initial dataset) and gives it a probability that the image is real. It consists of 3 layers of convolution layers and a dense layer. Both network are represented in Figure \ref{fig:dcgan}(a) and their details can be found in the supplementary information section.

\begin{figure}[htbp]
\centering
\includegraphics[width=160mm]{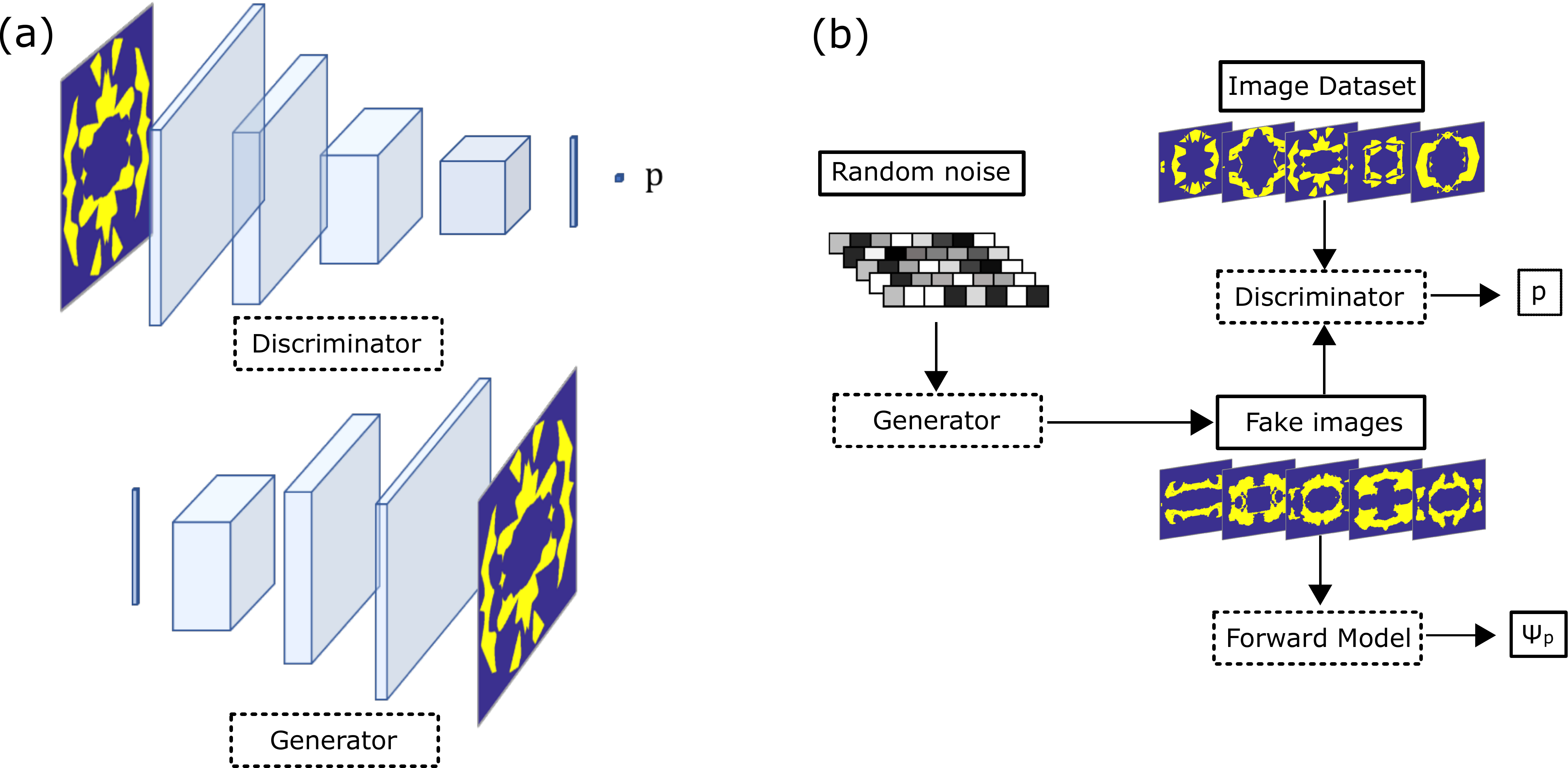}
\caption{(a) Discriminator and Generator networks. The Discriminator consists of 4 convolutions layers and one dense layer, with the output being the probability of the image being "real" (from the dataset) or "false" (from the generator). The Generator consists of 3 Transposed convolution layers and the output is a binary image of a shell configuration. (b) Block diagram of the DCGAN. The neural networks are represented with dashed lines, while data are represented with full lines. Output value $p(y)$ represents the probability of an image being tagged as real, while output $\Psi$ represents the scattering metric of the shell represented by the image. \label{fig:dcgan}}
\end{figure}

The adversarial play between the two networks is accomplished via their loss function, which is computed using the binary cross-entropy function :

\begin{equation}
    L(I) = -\frac{1}{N} \sum^{N}_{i}{y_i log(p(y_i))+(1-y_i) log(1-p(y_i))}
\end{equation}

where $I$ is the dataset of images, $N$ is the number of images in the dataset, $y_i$ is the label of the image (1 or 0), and $p(y_i)$ is the probability output. Loss of the discriminator is calculated using a sample of real and fake images with labels 1 and 0 respectively, while the loss of the generator is calculated using the sample of fake images only and using 1 as their label. This way, the discriminator is trained to differentiate the real and fake images, while the generator is trained to fool the discriminator by feeding it more and more realistic images which look like those of the dataset. 

To this branched network we add an additional path, which will calculate the scattering metric of the new fake images using the forward model. This way, we can add this value to the loss function of the generator so the new images not only aim to resemble the ones of the dataset but also to optimize cloaking. Schematic of the full DCGAN network is represented in Figure \ref{fig:dcgan}(b). The total loss function of the generator is thus given by :

\begin {equation}
L_t = \alpha_g L_g + \alpha_f L_f,
\label{eq:total_loss}
\end{equation}

where $L_t$ is the total loss, $L_g$ is the generator loss, $L_f$ is the forward model loss and $\alpha_g, \alpha_f$ are weighting coefficients. Care must be taken in order to choose weighting coefficients that balances cloaking optimization and generator efficiency.

One important detail of the previous network is the implementation of a rounding function on the fake images obtained from the DCGAN in order to have binary image input for the forward model. Since a rounding function's derivative is equal to zero, its direct use will lead to a vanishing gradient and the forward model wouldn't contribute to the training of the generator. We thus use a round function with a forced approximated derivative given by:
\begin{equation}
    df/dx = \sigma (1-\sigma) \qquad \qquad  \sigma = \frac{1}{1+e^{-10(x-0.5)}} 
\end{equation}

This function represents the derivative of a sigmoid function centered at 0.5, which is a close approximate of the rounding function.

\subsection{Feedback Loop Training}

DCGAN networks are effective at quickly generating and evaluating potential configurations for the cloacking shell. The addition of the forward network guides this generative process towards the optimization goal, in this case to minimize scattering. This method is limited though since the forward model isn't perfect at predicting the output to minimize, especially for configurations that differ considerably from those of the dataset. Some solutions might thus not be as optimal as the forward model predicts, and the model might also pass on a good configuration by wrongly predicting its output.

\begin{figure}[htbp]
\centering
\includegraphics[width=120mm]{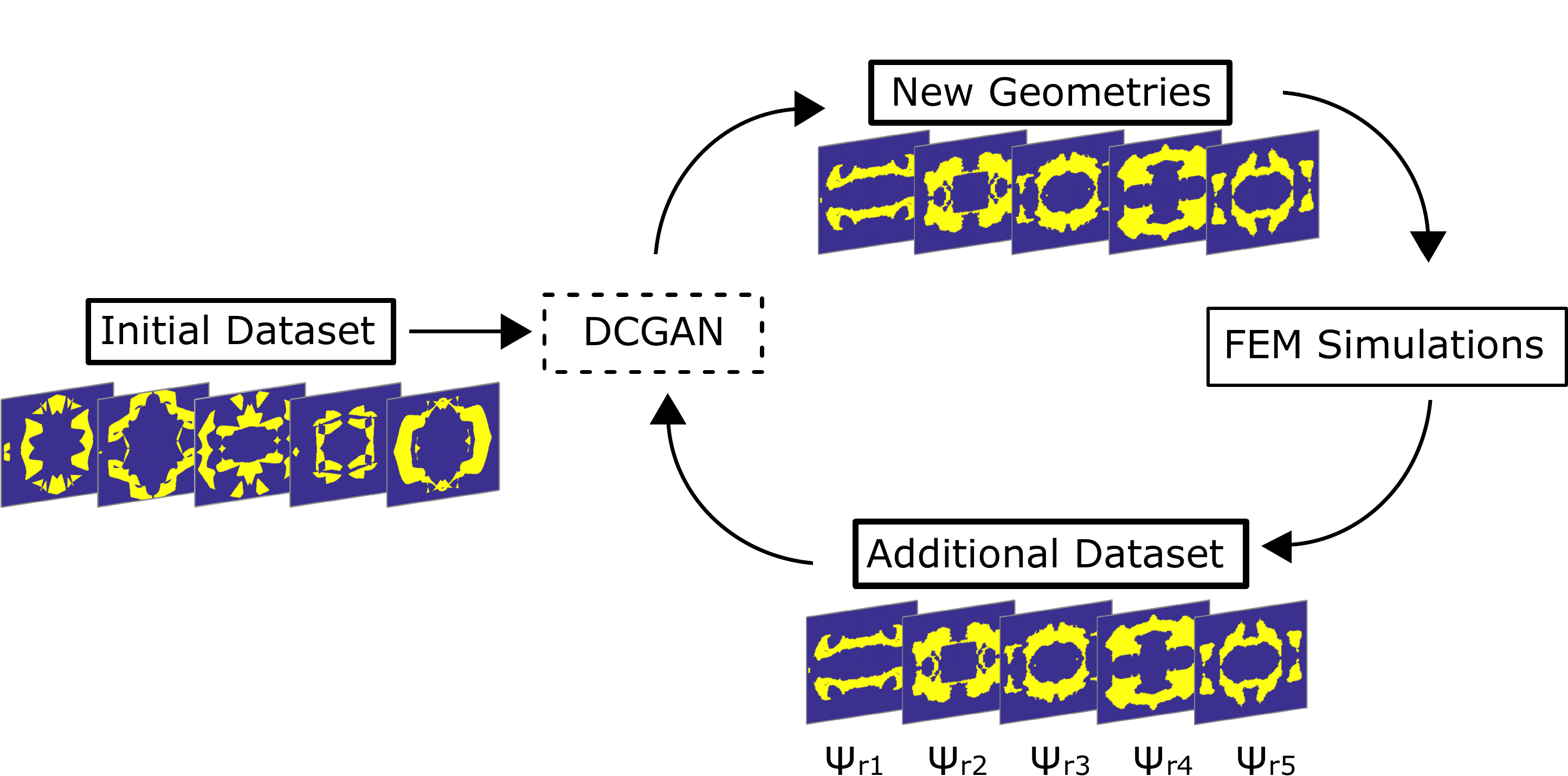}
\caption{Feedback loop training of the DCGAN. After each training of the DCGAN for 60 epochs, the 1000 best configurations are taken and simulated using FEM. The new ground truth data is used to retrain the forward model and the DCGAN is retrained with the new dataset. \label{fig:bloch}}
\end{figure}

For this reason, an iterative process of retraining the forward network is suggested for improving the solution search. The method is depicted in Figure \ref{fig:bloch}. The best solutions from the DCGAN according to the predicted scattering coefficient $\Psi_p$ are taken and simulated using FEM in order to obtain the actual scattering coefficient $\Psi_r$. Those solutions are then added to the dataset and the forward model is retrained. The new forward model is added in the DCGAN and a new solutions search is initiated, using the new and improved dataset.

This way, the forward model is improved by correcting any wrong prediction attributed to specific configurations which were deemed optimal. Furthermore, the dataset is improved by including good configurations, which will in turn lead the generator towards suggesting more favorable geometries.

\section{Results and discussion}

\subsection{Training of the DCGAN}

Training of the DCGAN usually requires a fine tuning of the parameters for each of the neural networks, whether it is the generator, the discriminator or the forward model. Both generator and discriminator are involved in a zero-sum game one with the other, since positive outcome for one means negative outcome for the other. They usually reach a Nash equilibrium and oscillate out of phase \cite{goodfellow2016nips}. The generator and discriminator need not to overpower one another, or else the generated configuration might be random and noisy, and as they differ considerably from those of the dataset the performance of the forward model will decrease significantly. 

\begin{figure}[htbp]
\centering
\includegraphics[width=160mm]{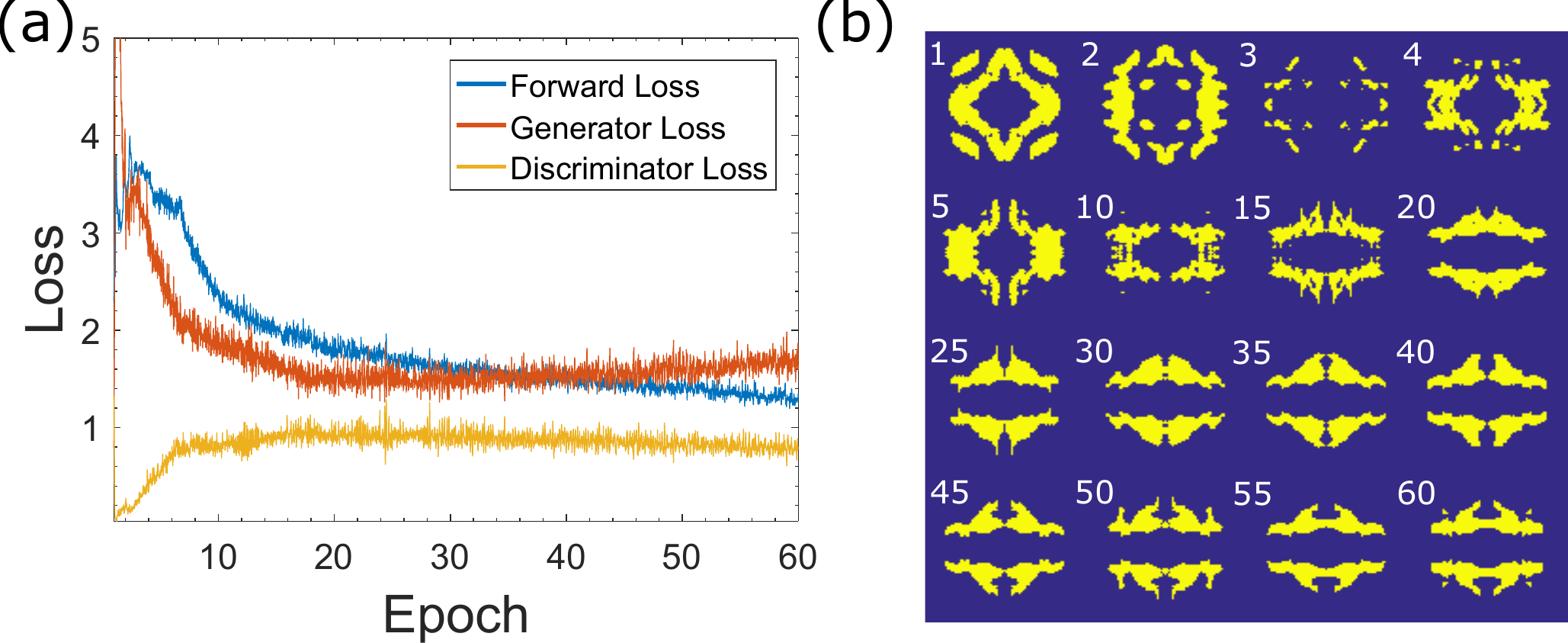}
\caption{(a) Forward, Generator and Discriminator Loss during training of the DCGAN. (b) Evolution of the generated images of the DCGAN for a specific noise vector during the training of the 5\textsuperscript{th} iteration of the feedback loop.\label{fig:epoch}}
\end{figure}

Adding an additional loss for the generator with the forward model creates a perturbation of this equilibrium, since the generator not only updates its weights to fool the discriminator but also to minimize scattering of the configuration. This leads, as can be seen in Figure \ref{fig:epoch}(a), to a generator reaching an equilibrium slightly above that of the discriminator, meaning that the generator might not perform fully at suggesting configuration resembling those of the dataset. Choosing the weighting parameters $\alpha_f, \alpha_g$ and  $\alpha_d$ of Equation (\ref{eq:total_loss}) carefully as to keep the difference between generator and discriminator loss as small as possible is thus crucial for obtaining adequate solutions. The coefficient $\alpha_g$ and $\alpha_d$ were set to 1 and the parameter $\alpha_f = 5 / <\Psi>$ for each generation of the method in order to adapt for lower loss with improved shell configuration.

Since the dataset doesn't contain similar type of images, the outputs of the generator remains dynamic and keeps evolving during the training even once the equilibrium between discriminator and generator has been reached. Figure \ref{fig:epoch} (b) plots an exemple of the output image for a specific noise vector during certain training epochs of the DCGAN (during 5\textsuperscript{th} iteration of the feedback loop). Even if the image converges towards a certain shape at epoch 20, little variations are still created for subsequent epochs. For this reason, a solution search is made after each epoch to maximize the number of different potential configurations.

\subsection{Feedback loop}

\begin{figure}[htbp]
\centering
\includegraphics[width=160mm]{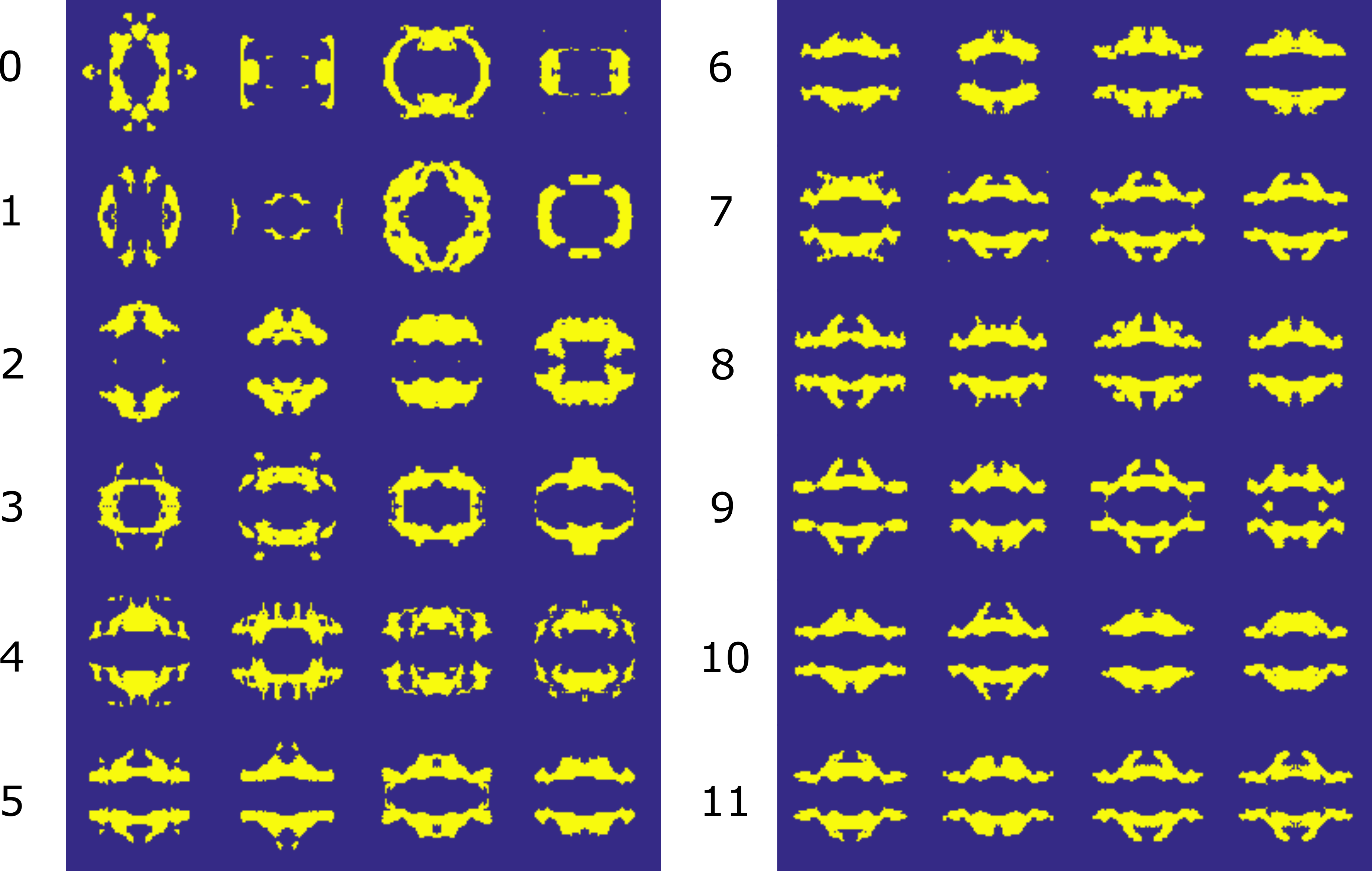}
\caption{Examples of 4 generated images at the end of the 60 epochs training of the DCGAN for each iteration of the forward loop.  \label{fig:gen_im}}
\end{figure}

In order to achieve optimization of the cloaking shell, 11 iterations of the training of the DCGAN were accomplished. In Figure \ref{fig:gen_im}, 4 random examples of the generator output are presented for each iteration of this feedback loop of the DCGAN. The proposed configurations are very diverse for the first few iterations, but converge towards one optimal configuration starting at iteration 5. This convergence is caused by two effects: on one hand, the forward model forces the generator to suggest optimal configurations, and on the other hand, the dataset is slowly getting more and more populated by those optimal configurations. In order to avoid getting the algorithm trapped in a local minimum, the $\alpha_f$ parameter of Equation (\ref{eq:total_loss}) needs to be chosen carefully low enough as to not collapse all the solutions of the generator towards the same configuration.

The strong point of the DCGAN is its ability to generate adequate configuration and to quickly predict the output of a certain configuration. The forward model takes about 0.1 ms for one prediction working on a Kaggle 17 GPU RAM kernel, while one COMSOL simulation takes 5-10s, making it roughly 50 000 - 100 000 times faster. We thus take advantage of this feature by testing 128000 configurations at each of the 60 epoch of the DCGAN.

\begin{figure}[htbp]
\centering
\includegraphics[width=160mm]{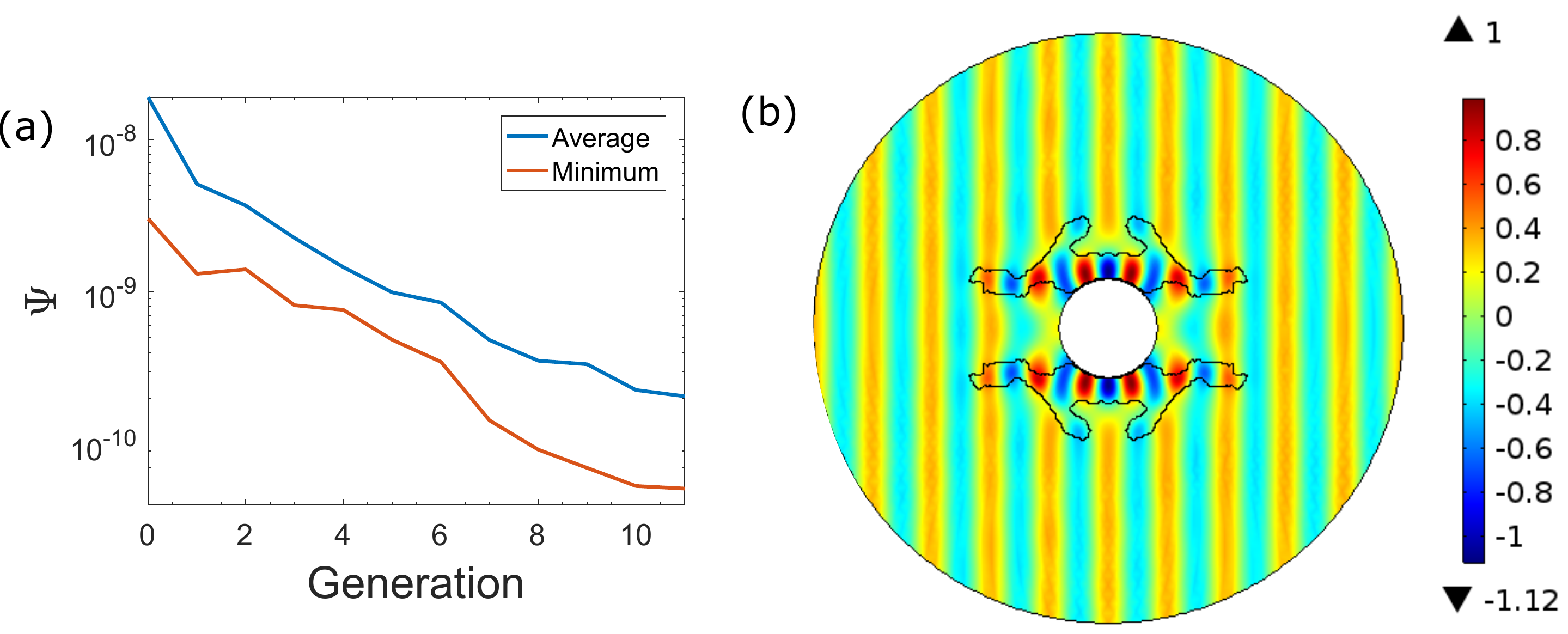}
\caption{Value of the average value (blue) and the minimum value (red) of $\Psi_r$ for the 1000 best configurations found by the DCGAN at every generation. The retraining of the DCGAN was done for 11 generations until the minimal value of $\Psi_r$ no longer improved. \label{fig:best_design} (b) Normalized value of the trasverse $H_z$ field with the optimal configuration.}
\end{figure}

After each training of the DCGAN the 1000 best configurations are taken according to their predicted value of $\Psi_p$ and are simulated in the FEM method in order to have their actual value $\Psi_r$. The loop is continued until no more improvements in the $\Psi_r$ is observed. The minimal value of $\Psi_r$ and the average value for each of the 11 iterations are plotted in Figure \ref{fig:best_design}(a). We can observe a continual improvement of the minimal value and the average value of $\Psi$ until it reaches $5.11 \times 10^{-11}$ W/m. This leads to a cloaking ratio of $0.0089$ when comparing to the value $5.77 \times 10^{-9}$ W/m without the cloak. This performance is comparable to solutions found using topology optimization \cite{andkjaer2012towards,andkjaer2011topology,fujii2013level} for similar dimensions of the shell and object compared to the wavelength. In Figure \ref{fig:best_design}(b) the total transverse magnetic field $H_z$ is plotted with the optimal cloak configuration found, which shows almost no distorsion of the wavefront compared to Figure \ref{fig:geo}(b). A strong concentration of the fields near the top and bottom regions of the object is observed, which is typical of the bending of the electromagnetic wavefront for cloaking shells.

\section{Conclusion}
In this paper we have demonstrated the use of deep learning for the optimization of an optical cloak. The suggested algorithm consists in a DCGAN architecture which is trained multiple times in a feedback loop in order to improve the solution search. The total scattered field coefficient, calculated with the Poynting vector at the outside boundary of the simulation domain of a FEM simulation, reached a ratio of 0.0089 of the field scattered without a cloak, which is comparable to results obtained using topology optimization.

\bibliographystyle{unsrt}  
\bibliography{references}  

\end{document}